\documentclass[conference]{IEEEtran}
\let\labelindent\relax
\usepackage{enumitem}

\makeatletter
\newcommand*{\rom}[1]{\expandafter\@slowromancap\romannumeral #1@}
\makeatother

\usepackage{color}

\usepackage{cite}
\usepackage[numbers]{natbib}

\ifCLASSINFOpdf
  \usepackage[pdftex]{graphicx}
  \graphicspath{{figures/}}
  \DeclareGraphicsExtensions{.pdf,.jpeg,.png,.eps,.tex}
\else
  \usepackage[dvips]{graphicx}
  \graphicspath{{figures/}}
  \DeclareGraphicsExtensions{.eps,.png}
\fi
\usepackage[labelformat=simple]{subcaption}

\usepackage[cmex10]{amsmath}
\usepackage{amssymb}
\usepackage{fixltx2e}

\DeclareMathOperator*{\argmax}{arg\,max}

\usepackage{algorithmic}

\usepackage{array}

\usepackage{mdwmath}
\usepackage{mdwtab}

\usepackage{eqparbox}

\usepackage[]{caption}
\usepackage{fixltx2e}

\usepackage{stfloats}

\usepackage{url}

\begin{document}
%
\title{Controlled Label Propagation: Preventing Over-Propagation through Gradual Expansion}

\author{

\IEEEauthorblockN{Aria Rezaei}
\IEEEauthorblockA{Department of Computer Engineering\\
Sharif University of Technology\\
Email: arezaei@ce.sharif.edu}
\and
\IEEEauthorblockN{Saeed Mahlouji Far}
\IEEEauthorblockA{Department of Computer Engineering\\
Sharif University of Technology\\
Email: mahloujifar@ce.sharif.edu}
\and
\IEEEauthorblockN{Mahdieh Soleymani}
\IEEEauthorblockA{Department of Computer Engineering\\
Sharif University of Technology\\
Email: soleymani@sharif.edu}
}


%


\maketitle

\begin{abstract}
Identifying communities has always been a fundamental task in analysis of complex networks. Many methods have been devised over the last decade for detection of communities. Amongst them, the label propagation algorithm brings great scalability together with high accuracy. However, it has one major flaw; when the community structure in the network is not clear enough, it will assign every node the same label, thus detecting the whole graph as one giant community. We have addressed this issue by setting a capacity for communities, starting from a small value and gradually increasing it over time. Preliminary results show that not only our extension improves the detection capability of classic label propagation algorithm when communities are not clearly detectable, but also improves the overall quality of the identified clusters in complex networks with a clear community structure.
\end{abstract}


%
\IEEEpeerreviewmaketitle

\section{Introduction}
Complex networks appear in a wide variety of domains. As a result, studying the structure of such networks has attracted a tremendous amount of attention throughout years. Real-world networks are usually comprised of \textit{communities} -- informally described as a group of nodes with a dense connection between themselves and a loose connection to the rest of the network.
As the building blocks of complex networks, communities reveals invaluable information about key features of the network. Retrieving the community structure can help us find the functional modules in biological networks, find groups of cohesive data cubes in large-scale databases, find groups of users in online social networks with similar attributes and interests, thus enabling us to develop effective marketing strategies in such networks, predict future interactions between users or study the emergence and popularity of ideas in social media\cite{spirin2003protein}\cite{krawczyk2009communities}\cite{backstrom2006group}. For a great survey refer to \cite{fortunato2010community}.

There is no single definition of communities. 
\citeauthor{radicchi2004defining} define communities in two senses\cite{radicchi2004defining}. Let node $v$ be in community $C$ with degree $d_v$, then the number of links between $v$ and other nodes within $C$ will be $d^{in}_v$, and the number of links between $v$ and the rest of the network will be $d^{out}_v$. A subgraph $V$ of a graph $G$ is a \textit{community in a strong sense} if:

\begin{equation} \label{eq:strongcom}
\forall v \in V: d^{in}_v > d^{out}_v
\end{equation}

Similarly a subgraph $V$ of graph $G$ is a \textit{community in a weak sense} if:
\begin{equation} \label{eq:weakcom}
\sum\limits_{v \in V} d^{in}_v > \sum\limits_{v \in V} d^{out}_v
\end{equation}

There are various measures quantifying the quality of clusters. One of the most popular measures is \textit{modularity}\cite{newman2006modularity}. Knowing that real-world networks possess strong community structure compared to random networks, modularity measures the difference between a given partitioning in a certain graph and the same partitioning in a random graph with the same distribution of degrees. Modularity in a partitioning with $C$ clusters can be written as below:

\begin{equation} \label{eq:mod}
Q = \sum\limits_{{c_i}\in C}\Big[\frac{m_i}{M}-(\frac{d_i}{2M})^2\Big]
\end{equation}
Where $m_i$ is the number of edges inside partition $c_i$, $d_i$ is the total degree of nodes inside $c_i$, $C$ is the set of all clusters and $M$ is the total number of edges.

The growing need for retrieving communities from complex networks in addition to rapid growth of the size of data, highlights the importance of scalable and accurate methods to detect communities in such networks. With modularity as objective function, graph partitioning can be seen as an optimization problem. Many algorithms for community detection have been proposed over the last decade, setting modularity optimization as their ultimate goal, which has proved in practice to retrieve communities of great quality\cite{louvain}\cite{cnm}\cite{extrememodularity}. There are a class of algorithms exploiting the power of linear algebra to detect communities using the eigenvectors of the Laplacian matrix\cite{1973spectral}\cite{1990spectral}. Another class of algorithms use the fact that clusters have weak connection between themselves; knowing this, this class of algorithms find minimal cuts and divide the graph recursively, leading to a dendrogram of node-cluster membership\cite{metis}\cite{flowmqi}. There are also novel approaches that do not completely fit in the previous categories which are built around statistical and mechanical phenomena in real world\cite{spinglass}\cite{affinitypropagation}\cite{pottsmodel}.

All these methods try to optimize a global objective function or use the whole structure of a network to divide it into clusters. There is a problem with this kind of approach, especially encountered in social networks; individuals in such networks does not join communities to increase a global quality function, they rather join them to improve their own utility function, be it more enjoyment through joining a group of people with similar interests in Facebook, or following a politician in Twitter in order to stay in touch with the latest news in politics\cite{chen2010game}. Apart from that, it is proved that algorithms based solely on modularity optimization fails to detect communities with small size as the size of the network increases, which is famously known as \textit{resolution limit} of modularity\cite{fortunato2007resolution}. 
All this evidence lead us to choosing a \textbf{node-centred} approach. As an example of node-centred approaches, some algorithms employ a game-theoretic approach to find communities. Let nodes be agents in a game with a personal utility function. A strategy in this game which leads to a nash or local equilibrium will yield a community structure in the network\cite{chen2010game}. Another approach with near linear time is \textit{Label Propagation Algorithm}(LPA)\cite{raghavanlabel}. In this method, every node is assigned a unique label in the initial condition of the network. Afterwards, in each iteration, nodes are traversed using a random order and every node acquires the label which is most frequent among its neighbors. This method has one serious problem though; in some networks, one of the labels will \textit{over-propagate} and sweep through other labels, leading to detection of the whole network as one giant community, which renders this method counterproductive in dense networks with unclear community structure. This phenomenon is often called \textit{flood-fills}. The focus of this article is overcoming this flaw and improving the accuracy of LPA when community structure is hardly detectable.

The rest of the article is structured as follows. Section \rom{2} reviews potentials and flaws of LPA along with subsequent extensions to overcome its flaws. Section \rom{3} presents the formal definition of \textit{Controlled Label Propagation Algorithm} using gradual expansion of communities and discusses the rationale behind choosing this approach. In Section \rom{4} we empirically compare our method with LPA and two of its extensions along with several state-of-the-art algorithms in real-world networks in addition to standard network benchmarks. Lastly, section \rom{5} discusses further research related to our work in addition to a conclusion.

\section{Background and Motivation}

In this section we will discuss the considerable potentials in LPA and why it is important to improve this method for further uses. We will also cover some of the flaws of this algorithm and previous efforts do deal with these flaws.

\subsection{Where label propagation prevails?}

One of the biggest advantages of LPA is its time-efficiency. Raghavan et al. state that 95\% of the nodes reach their final state in 5 iterations\cite{raghavanlabel}. Although the experiments by various authors suggest that the number of iterations needed to reach equilibrium state, $I$, grows very slowly with the size of the network, it is not fully understood yet. \citeauthor{leung} have created a class of networks on which $I$ increase logarithmically with the size of the network\cite{leung}. Although this class of networks is highly unlikely to appear in natural graphs, it shows that the efforts to find an upper-bound for $I$ has gone no further than $\log{N}$ in a network with size $N$. Since the time complexity of LPA is $O(MI)$ in a graph with $M$ edges, in the worst case, its time complexity will be $O(M\log{N})$ which is still considerably fast compared to other methods.

Apart from its time-efficiency, LPA can be trivially extended to directed and weighted graphs with negative links. It also does not need to know the number of communities \textit{a priori}, so in domains with zero knowledge about the network, it can be quite useful.

Leung et al. believe that since every node only requires information from its neighbors, LPA can be easily run in a distributed environment in almost constant time\cite{leung}. This is particularly important with the emergence of ubiquitous computing and mobile social networks. All the computation will be distributed into nodes and since every node has very few neighbors compared to the size of the graph, the community structure of large-scale networks can be revealed in little time. The authors also suggest that since every node updates its state from \textit{current} formation of its neighbors, LPA can be used to detect communities in a dynamic environments when nodes and links might come and go\cite{leung}. Furthermore, nowadays there is growing concern regarding the privacy of information in social networks. Keep in mind that every node only receives information from nodes he is already in contact with. In addition, the topology of the whole graph is never revealed to the nodes. As a result, LPA also guarantees an acceptable level of privacy when run on social networks.

Moreover, the approach adopted by LPA algorithm is completely intuitive regarding real world communities. Consider the following scenario; there are $N$ people invited to a ceremony. Assuming friendship to be a \textit{zero-one} relationship, in which \textit{zero} means no friendship and \textit{one} means friendship, their friendship status forms an undirected and unweighted graph. After the ceremony starts, people will try to form \textit{circles} with their friends in order to be near them and enjoy their companionship. If we rearrange the circles with the communities retrieved from their friendship graph using LPA, the new circles will remain the same throughout the ceremony, since no one can join a \textit{better} circle. In other words, there is no one dissatisfied by her community and these communities provide an equilibrium state in real life. Formally, we can call a node $v$ in community $C_i$ \textit{dissatisfied} if there exists another community $C_j$ such that $d^{C_j}_v > d^{C_i}_v$, where $d^{C}_v$ is the number of links from $v$ to nodes within $C$.
As an inherent characteristic of LPA, there are no dissatisfied nodes in the retrieved communities. However, this does not hold for many other algorithms. We have tested several algorithms, namely multilevel modularity optimization of Blondel et al.\cite{louvain}, greedy modularity optimization of Claust et al.\cite{cnm} and random walk community detection of Rosvall et al.\cite{infomap}, to find the percentage of dissatisfied nodes in a network given a community structure by these algorithms [Results are omitted]. Although our experiments did not show a considerable portion of nodes being dissatisfied (on average, less than 5\% in five collaboration networks retrieved from \cite{snapnets}), keep in mind that this tiny portion of nodes can be considered as \textit{false positives} in domains where it is crucial that nodes belong to the optimal community in their own perspective, like in social networks.

\subsection{Where label propagation fails?}

As we mentioned before, on certain graphs, LPA fails to detect the community structure of the graph and reports the whole graph as one community. This is apparently because the speed of formation for different communities in the network varies significantly. In other words, the core of stronger communities are formed in the early stages, while weaker communities has not yet reached consensus. Furthermore, when a community has not reached its final state, it is comprised of several smaller pieces with different labels, possessing a weak core which is vulnerable to propagation of foreign labels. In a nutshell, stronger communities, exploiting the lack of unity in weaker communities, often sweep through the cores of small pieces of weaker communities and attract all of their members. This usually leads into one giant community and several smaller ones. On rare cases, the situation is exacerbated when there is one label strong enough to overwhelm all other labels, leaving every node with the same label in the end.

Leung et al. have addressed this issue by employing a technique called \textit{Hop Attenuation}\cite{leung}. This technique is based on the observation that the diameter of communities ought to be tiny proportional to that of the whole graph. With hop attenuation, the labels lose their strength while they propagate. Meaning that after traversing away from its center, the label's strength is exhausted and will ultimately vanish. This phenomenon can be formally described as below:

\begin{equation}\label{eq:hopatt}
S_v(\mathcal{L}) = \max(S_u(\mathcal{L}):u \in \mathcal{N}(v)) - \delta
\end{equation}

Where $S_v(\mathcal{L})$ is the strength of label $\mathcal{L}$ when propagated by node $v$, $\mathcal{N}(v)$ is the neighborhood set of node $v$ and $\delta$ is a parameter used to decrease the strength of labels when they traverse a single link. The authors realized that choosing a constant $\delta$ irrespective of the network may lead to a substantial decrease in the quality of communities. Knowing this, they proposed some methods to adaptively change $\delta$ through either using the current number of iterations or simply know it \textit{a priori} as a parameter given to the algorithm.

\citeauthor{leung} used another powerful heuristic to avoid imbalanced propagation of the labels called \textit{Node Preference}. They realized that blindly valuing every neighbor of a node the same might be an apparent reason for the over-propagation phenomenon. Given that finding maximal label in LPA is simply finding the label with maximum occurrence among neighbors, Leung et al. changed the formulation to the following:

\begin{equation}\label{leunglabel}
\mathcal{L}^{i+1}_v = \argmax_{\mathcal{L}^i}\sum\limits_{u\in \mathcal{N}(v)} S_u(\mathcal{L}^i_u)\dot{f(u)^m}
\end{equation}

Where $\mathcal{L}^i_v$ is the label assigned to node $v$ in iteration $i$, $S_v(\mathcal{L}^i_v)$ is the strength of $v$'s label in iteration $i$, $\mathcal{N}(v)$ is the neighborhood set of $v$ and $f(v)$ is any comparable function on nodes, such as betweenness centrality\cite{verbetweenness}, degree centrality or any other measure.

\citeauthor{defensiveoffensive} have taken advantage of both \textit{node preference} and \textit{hop attenuation} and proposed two new strategies called \textit{defensive preservation} and \textit{offensive expansion} of communities\cite{defensiveoffensive}. In defensive preservation, the preference is given to nodes in the core of communities. On the other hand, in offensive expansion, the preference is given to nodes in the border of communities. The core and the border are identified using a simple random walk algorithm. They have devised two algorithms, namely \textit{DDALPA} and \textit{ODALPA}, for defensive and offensive strategies respectively. The authors show that DDALPA results in high recall, whereas ODALPA gives high precision. Furthermore, in an attempt to take advantage of merits in both DDALPA and ODALPA, they have created a method called \textit{BDPA} which simply runs them one after another to find communities of good quality in small-size networks.
Finally, exploiting the \textit{core-periphery} structure of large-scale networks\cite{leskovec2009coreperiphery}, they developed a hierarchical algorithm, called \textit{DPA}, that has the efficiency of BDPA in addition to recursively finding communities in the giant core of the graph. Going into the details of these algorithms is out of the scope of this article, it suffices to say that the results of both algorithms, especially DPA, on large-scale graphs is comparable to state-of-the-art algorithms in terms of modularity measure.

\citeauthor{lpaconstraints} took a different path. They realize that over-propagation stems from the objective function of LPA\cite{lpaconstraints}. Let $M$ be a matrix of $N$ rows and $C$ columns ($C$ is the number of communities), then $M_{ij}$ is one, if node $i$ has label $j$, or zero otherwise. Each row in this matrix is the \textit{membership vector} of a node. When a node is updated, it adopts the most frequent label among its neighbors, thus increasing the sum of dot products between its own membership vector and that of its neighbors. Formally, when we update node $v$, we increase:

\begin{equation}\label{eq:lpaobj1}
\sum\limits_{u\in \mathcal{N}(v)}\sum\limits_{i=1}^C M_{vi}M_{ui}
\end{equation}

Applying adjacency matrix, $A$, to (\ref{eq:lpaobj1}) yields:

\begin{equation}\label{eq:lpaobj2}
\sum\limits_{i=1}^N \sum\limits_{j=1}^C M_{vj}M_{ij}A_{vi}
\end{equation}

Finally, since LPA does this for every node in an iteration, the overall objective function, $H$, of LPA becomes:

\begin{equation}\label{eq:finallpaobj}
H=\sum\limits_{i=1}^N \sum\limits_{j=1}^N \sum\limits_{k=1}^C M_{ik}M_{jk}A_{ij}
\end{equation}

With a little thought, it will become clear that LPA is simply increasing the number of edges connecting two nodes within the same community. Thus, the algorithm reaches its global maximum when every node is assigned the same label. To avoid this undesirable result, \citeauthor{lpaconstraints} changed the objective function of LPA to below:

\begin{equation}\label{eq:changedlpa}
H'=H-\lambda G
\end{equation}

Where $H$ is the previously mentioned objective function of LPA and $G$ is a penalty function that diverts the algorithm from its previous undesired global maximum with a coefficient $\lambda$. \citeauthor{lpaconstraints} have proposed several $G$ functions, two of them changes $H$ into modularity of unipartite and bipartite networks\cite{lpaconstraints}, thus enabling LPA to locally maximize modularity. This is a very impressive approach since it keeps the potential benefits of LPA (parallelism, privacy insurance, etc.) while overcoming its flaws and retrieving communities of excellent quality.

We now look at the flood-fill phenomenon from a different angle. LPA has two modes, \textit{asynchronous} and \textit{synchronous}\cite{raghavanlabel}. They both provide similar results, but the synchrobous mode gives a better insight about the underlying reasons of flood-fills. At the first iteration of LPA, every node is assigned a unique label, thus every label in the neighborhood of a node $v$ has the same frequency. As a result, the probability of a node $v$ choosing a label $\mathcal{L}$ in its neighborhood is $\frac{1}{d_{v}}$ where $d_{v}$ is the degree of node $v$. We can easily compute the expected number of nodes in a label after the first iteration as below:

\begin{equation}\label{eq:estdist}
\mathbb{E}[N(\mathcal{L}_u)] = \sum\limits_{v\in \mathcal{N}(u)} \frac{1}{d_{v}}
\end{equation}

Where $\mathcal{L}_u$ is the label first assigned to node $u$. We call the expected number of nodes choosing a label $\mathcal{L}_v$ after first iteration \textit{attraction power} of node $v$ and denote it by $\mathcal{A}(v)$.

\begin{figure}[t!]
	\small
	\begin{center}
		\input{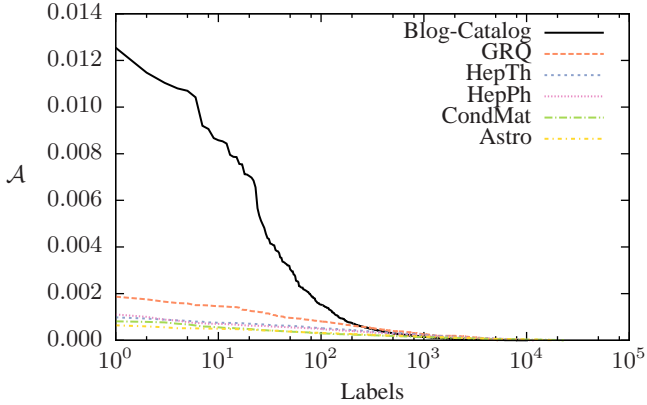}
	\end{center}
	\caption{\label{fig:attention}(Color online) We have calculated the attraction power ($\mathcal{A}$) of labels in the six networks, which is an estimation of the number of nodes holding each label after the first iteration of LPA. As the plot suggests, attraction power is much more equally spread across nodes in the five co-authorship graphs compared to the blog-catalog network. The results show that it might be too late to prevent a label from flooding the whole graph even after the first iteration. It also shows that we can determine how likely LPA is to fail based solely on the structure of the graph}
\end{figure}
We have calculated $\mathcal{A}$ for each label after the first iteration of LPA on five co-authorship networks\cite{snapnets}, as examples of sparse networks with clear community structure, and a friendship network in blog-catalog\cite{asunets} as an example of dense social network which LPA fails to retrieve its community structure. We have plotted the results, with nodes sorted descending by $\mathcal{A}$, in Fig.\ref{fig:attention}. As the plot suggests, the decrease in blog-catalog network is far more rapid compared to five co-authorship networks. The distribution of $\mathcal{A}$ is important for two reasons. First, it enables us to anticipate the failure of LPA in networks before the algorithm starts. Second, Fig.\ref{fig:attention} shows that imbalanced growth of communities starts from the very first iteration, meaning that any strategy to control over-propagation of labels based on the results of previous iterations may fail. This highlights the importance of monitoring growth of communities from the very first iteration. In the next section we propose a simple yet effective way to prevent over-propagation starting from the first iteration.

\section{Controlled Label Propagation}

Before we go on and talk about preventing over-propagation, there is another problem with LPA that needs to be deal with. In LPA, when there are other labels in the neighborhood with the same frequency as a node's current label, the existing label will not change. Meaning that if a bad decision is made in the early stages, there is no come-back mechanism to escape from it. In terms of objective function of LPA, there is no mechanism to escape from an early local maximum and we are satisfied with with the first local maximum we reach. There has been a technique proposed by \citeauthor{lpaconstraints} that randomly walks on the local maximum to escape from it\cite{lpaconstraints}. This might be dangerous as it might randomly change the label of several border nodes, which decreases the defensive capabilities of a community and might lead to an \textit{avalanche effect} in which nodes' labels are changed one after another because of the change in border nodes' labels, which leads to a drastic change in the community structure after some iterations. To prevent this, we have acquired a strategy inspired by \textit{Simulated Annealing}\cite{simul} in which nodes tend to change a \textit{good} label -- a label with maximal frequency among the nodes' neighbors -- for another good label in the early stages. However they lose this tendency over time, meaning that they will hold their good label if there are no better labels in the final stages, since we do not want established cores to disappear. This can be done by a decreasing probability function $p(t)$, starting from $1$ and ending in $0$. If $p(t)$ holds, we randomly choose a label among good labels; if not, we will hold the current label if there are no better labels.

Now we can devise a method to prevent all nodes from ending up with the same label. We previously mentioned that the main reason behind flood-fills is the rapid formation of the core of some communities and sluggish growth of others. So the key to this problem is ensuring "fair" growth of communities in each iteration. To ensure that weaker communities have a chance to grow, we put a capacity for all communities as a function $\mathcal{C}(t)$ where $t$ is the current iteration. When a label's population reaches the capacity, it can no longer attract new nodes. We define $\mathcal{C}$ is follows:

\begin{equation}
\mathcal{C}(t) = \Big( \Big[\frac{kt}{T}\Big] + 1\Big)\times\frac{N}{k}
\end{equation}

Where $k$ is the number of times we increase the capacity of communities (the maximum number of nodes a label can be assigned to), $t$ is the current number of iteration, $T$ is the maximum number of iterations and $N$ is the size of network. It is clear that this function increases the capacity every $\frac{T}{k}^{th}$ iteration by $\frac{N}{k}$, starting from $\frac{N}{k}$ and ending in $N$. As a result, in the final stages there are actually no constraint on the size of communities. Furthermore, We call the iterations between two increases in capacity a \textit{cycle}, so $k$ will be the number of cycles.

The rationale behind our strategy to stop popular labels from attracting new nodes is in fact giving a chance to weaker communities to form a core. In this way, a node $v$ might choose a popular label $\mathcal{L}$ when it is not full. During the rest of the cycle, a number of nodes in $v$'s neighborhood find $\mathcal{L}$ full and join a sub-optimal label $\mathcal{L}'$. With enough nodes joining $\mathcal{L}'$, $v$ might change its mind and join $\mathcal{L}'$ during the later iterations of the current cycle and form the core of a new community along with a number of its neighbors. The cores might not be of good quality in the early cycles, but when we reach a new cycle and increase $\mathcal{C}$, those nodes who are not content with their current label will join their desired label before it gets full again. However, the core nodes of a community who have acquired a label will keep their label and attract new nodes. In this way we are helping weaker communities to form their cores while still giving nodes complete freedom in the final cycles to choose their desired label. We call our method Controlled Label Propagation Algorithm (CLPA).

In the next section we show the result of our tests and prove that our method is more effective than previously proposed extensions to prevent flood-fills. Furthermore, the overall quality of clusters are also increased on the various datasets we tested our method on.

\section{Experiments}

\subsection{Sparse Networks}
We have tested CPLA against LPA on a wide variety of networks. Although increasing the quality of clusters was not our primary intention, the results reveal that on certain graphs, the modularity score of clusters yielded by CLPA is on average much higher than LPA. The data sets used for this part include five co-authorship networks\cite{snapnets}, namely GRQ, HepTh, HepPh, CondMat and Astro, retrieved from articles on different topics in Physics (for more information about these networks refer to \cite{snapnets}). There are two location-based online social networks\cite{snapnets}, namely Gowalla and Brightkite and an e-mail client network named Enron\cite{snapnets}. There is also a graph of a piece of Youtube's social network, along with a Computer Science co-authorship network of DBLP and product co-purchased network of Amazon\cite{snapnets}. We ran our algorithm with three different \textit{k}s, 50, 100 and 200, along with LPA on all the networks for fifty to two times, depending on size of the networks. Table.\ref{tab:mod} contains the name of the network, the number of nodes $N$, the number of edges $M$ and the modularity score $Q$ achieved by two algorithms. We have used the igraph python library to compute modularity for retrieved communities\cite{igraph}.

\begin{table}

	\centering
	\caption{Modularity scores for eleven data sets averaged on 50 to 2 realizations based on the size of networks. We have also shown the variance of $\mathcal{A}$ as an indicator of unfairness of attraction power in the networks.}
	\label{tab:mod}
	\begin{tabular}{|c||c||c||c||c||c|}
	\hline
	Name & \textit{N} & \textit{M} & $var(\mathcal{A})$ & \textit{CLPA} & \textit{LPA} \\
	\hline
	GRQ & 5.24K	& 14.48K &	1.0 &\textbf{0.797} & 0.735\\
	\hline
	HepTh & 9.88K &	25.97K & 1.1 & \textbf{0.671} & 0.627\\
	\hline
	HepPh & 12.01K & 0.12M	& 1.2 & \textbf{0.497} & 0.488\\
	\hline
	CondMat & 12.01K & 0.12M & 1.3 & \textbf{0.633} & 0.578\\
	\hline
	Astro & 18.77K & 0.20M & 1.1 & \textbf{0.450} & 0.323\\
	\hline
	Enron & 36.69K & 0.18M & 80.8 & \textbf{0.473} & 0.338\\
	\hline
	Brightkite & 58.23K & 0.21M & 6.9 & \textbf{0.623} & 0.557\\
	\hline
	Gowalla & 0.20M & 0.95M & 102.4 & \textbf{0.618} & 0.503\\
	\hline
	DBLP & 0.32M & 1.05M & 2.4 & \textbf{0.697} & 0.622\\
	\hline
	Amazon & 0.33M & 0.93M & 2.1 & \textbf{0.786} & 0.709\\
	\hline
	Youtube & 1.13M & 2.99M & 233.9 & \textbf{0.682} & 0.555\\
	\hline
	\end{tabular}
\end{table}

\begin{table*}[!t]
	\renewcommand{\arraystretch}{1.5}
	\caption{We tested the Blog-Catalog network using the algorithms shown in the table. We have shown the average NMI over ten realization. As the results reveal, LPA and both of its extensions, BPA and DPA, fail to capture any meaningful community strucutre in this network. This highlights the importance of efficient ways to prevent the flood-fill phenomenon. Also worth noticing is the failure of the Louvain algorithm to capture the essence of ground-truth communities, resulting in a \textit{zero} NMI over ten realization. Finally, note that CNM yields better results than ours with considerably higher execution time, around 50s in a network of this size, which might cause troubles if one intends to use CNM  on massive graphs.}
	\label{tab:nmi_bc}
	\begin{center}
		\begin{tabular}{|c||c||c||c||c||c||c||c||c||c||c|}
		\hline
		$N$ & $M$ & $\bar{d}$ & $d_{max}$ & \textit{CLPA} & \textit{LPA} & \textit{DPA} & \textit{BPA} & \textit{Infomap} & \textit{CNM} & \textit{Louvain} \\
		\hline
		10.3K & 333.9K & 64.7 & 3992 & 0.0006 & {\color{red}0.0} & {\color{red}0.0} & {\color{red}0.0} & 0.0003 & 0.0009 & {\color{red}0.0} \\
		\hline
		\end{tabular}
	\end{center}
\end{table*}

Table.\ref{tab:mod} reveals that letting weaker communities grow improves the overall quality of detected communities. In section \rom{2}.B we conjectured that the failure of LPA stems from unfair distribution of attraction power ($\mathcal{A}$). In order to measure unfairness of $\mathcal{A}$, we have calculated the variance of this value for all nodes in each of the networks and shown the results in Table.\ref{tab:mod}. Note that out of five networks which yielded the highest percentage of increase in modularity, four are also in the top five networks with highest variance in $\mathcal{A}$. Also, the five networks with the lowest increase in modularity are in the bottom six networks in regard to their variance in $\mathcal{A}$. The only anomaly is \textit{Astro}, which is the $9^{th}$ network in respect to $var(\mathcal{A})$, but yields the second highest increase in modularity. This shows that the failure of LPA is not solely related to unfair wiring of the links in the networks. This calls for an in-depth analysis to broaden our understanding of the underlying reasons of LPA shortcoming in different networks.

\subsection{Dense Networks}
As an example of social networks with dense structure in which LPA fails completely to retrieve any community structure, we have found the Blog-Catalog social graph which we formerly analyzed in section \rom{2}. In this network, the average and maximum degree (shown in Table.\ref{tab:nmi_bc}) is significantly higher than previous networks we tested in section \rom{4}.b. The node with the maximum degree is connected to 38.7\% of the nodes. Furthermore, there are 62 nodes connected to more than 10\% of the network. The symptoms we mentioned are quite rare in social network structures. However, this shows that there exists instances of social networks, although rare, in which LPA fails to retrieve any meaningful set of communities. This means that we have legitimate concerns to stop this phenomenon.

The network comes with a set of ground-truth communities, which enables us to measure the similarity between retrieved communities with the true ones. One of the popular similarity measures is Normalized Mutual Information (NMI). Due to the overlapping nature of the ground-truth communities, we have used the code provided by \citeauthor{overlappingnmi} to calculate NMI between our communities and the true ones\cite{overlappingnmi}. Note that since ground-truth communities significantly overlap, the NMI score will be very low in general for any graph partitioning method, but it is still useful for the sake of comparison.

We have compared our algorithm's performance to the state-of-the-art algorithms, such as the multi-level modularity optimization algorithm of \citeauthor{louvain} (denoted by Louvain\cite{louvain}), the greedy modularity optimization of \citeauthor{cnm} (denoted by CNM\cite{cnm}) and the famous infomap algorithm\cite{infomap} proposed by \citeauthor{infomap}. We have also tested the data set with two of LPA's most successful extensions, namely BPA and DPA, proposed by \citeauthor{defensiveoffensive}\cite{defensiveoffensive}\cite{bpa}. The results, shown in Table.\ref{tab:nmi_bc}, show that not only our algorithm prevents a flood-fill and captures a community structure in this network, but also our communities yield relatively good results compared to other methods, except for CNM which is considerably slower than others. We believe that the relatively good resemblance between our communities and the true ones originates from the nature of LPA algorithm, which is completely intuitive regarding how social networks are formed in real world. Meaning that if one incorporates a strategy to ensure the balanced growth of communities with LPA's basic method of community detection, the results can be of great quality.

\subsection{Networks with planted partitions}
\begin{figure*}[!tbh]
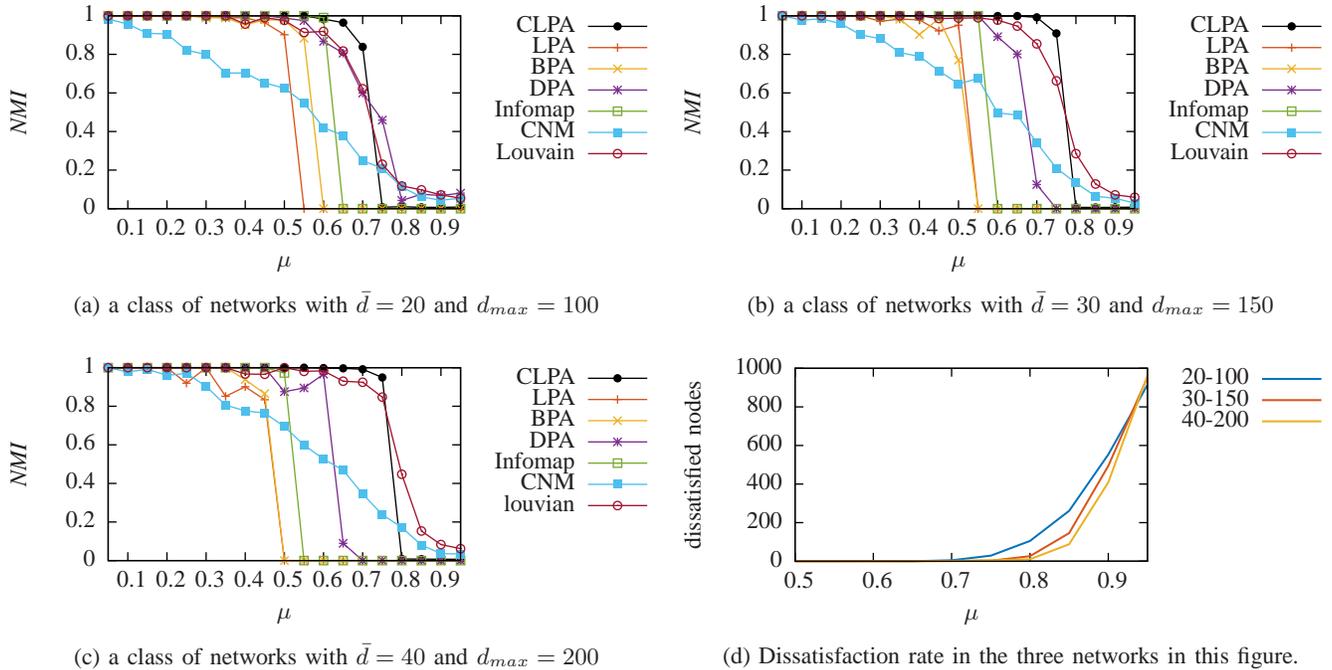

	\centering
	\begin{subfigure}{0.48\linewidth}
		\small
		\input{figures/nmi_20_100.tex}
		\caption{a class of networks with $\bar{d}=20$ and $d_{max}=100$}
		\label{fig:nmi:20-100}
	\end{subfigure}	
	\begin{subfigure}{0.48\linewidth}
		\small
		\input{figures/nmi_30_150.tex}
		\caption{a class of networks with $\bar{d}=30$ and $d_{max}=150$}
		\label{fig:nmi:30-150}
	\end{subfigure}
	\begin{subfigure}{0.48\linewidth}
		\small
		\input{figures/nmi_40_200.tex}
		\caption{a class of networks with $\bar{d}=40$ and $d_{max}=200$}
		\label{fig:nmi:40-200}
	\end{subfigure}	
	\begin{subfigure}{0.48\linewidth}
		\small
		\input{figures/unhappy.tex}
		\caption{Dissatisfaction rate in the three networks in this figure.}
		\label{fig:nmi:unhappy}
	\end{subfigure}
	\caption{
(Color online) We have tested several algorithms on \citeauthor{lancichinetti2008benchmark} benchmarks\cite{lancichinetti2008benchmark}, using different $\bar{d}$ and $d_{max}$. As the figures suggest, our algorithm surpass others by a big margin as the density ($\bar{d}$) increases. In addition, on the bottom-right figure, we have shown that the number of dissatisfied nodes increases dramatically as $\mu$ grows, meaning that without significantly changing the core of LPA, it is extremely difficult to detect communities in environments with very high $\mu$.
	}
	\label{fig:nmi}
\end{figure*}

In order to realize what parameters are responsible for the failure of LPA and get a better understanding of behavior of different algorithms in networks with a vague community structure, we have used \citeauthor{lancichinetti2008benchmark} benchmark\cite{lancichinetti2008benchmark}. To challenge the detection power of these algorithms, we have set the size of the network ($N$) constant, then created networks of different densities ($\bar{d}$) and community structure clarity (mixing parameter or $\mu$). In order to study the effects of fairness of degree distribution, we have also used networks with different maximum degree ($d_{max}$). The results are depicted in Fig.\ref{fig:nmi:20-100}, Fig.\ref{fig:nmi:30-150} and Fig.\ref{fig:nmi:40-200}. As you can see, the algorithms based on propagation of data, including LPA and all of its extensions along with Infomap, follow a trend of giving high quality results and then abruptly falling to zero. The exception is the DPA algorithm for which the decrease in the quality of clusters start sooner than others, but it keeps giving mediocre clusterings before completely failing to identify any useful structure out of the networks.
The pattern of behavior is different for two modularity optimization algorithms, namely Louvain and CNM, as they tend to have a more \textit{smooth} fall. However, the smoothness is both good and bad. Good, because in networks with a vague community structure, they continue to yield communities with some degree of quality. It is also bad, since the decrease in the quality of clusters, compared to our algorithm, starts sooner. In a nutshell, even though CNM and Louvain keep on yielding non-zero NMI on later stages and hit the ground slower than us, but they give sub-optimal solutions when our algorithm results in optimal communities. All in all, our algorithm keeps giving near-perfect results, while other results suffer from an amount of inaccuracy, rendering us superior in graphs with unclear community structure. Keep in mind that we are achieving better results in higher $\mu$ while giving perfect results on lower ones. Meaning that we are not sacrificing accuracy on the start of $\mu$ spectrum to attain higher accuracy in the end. Furthermore, even though other algorithms might experience sporadic decrease in NMI even before $\mu=0.5$, our algorithm stays firmly on top of the chart and yield near-perfect results in any case.

Looking at the \textit{landing points} -- the point in which NMI reaches zero -- of the algorithms can give us a great insight on the impact of the density of the networks on detection power of these algorithms. There is one group of algorithms, including LPA, BPA, DPA and Infomap, that lose their detection power as the density of the network increases. In other words, the landing point of these algorithms move to the left as the density grows. Interestingly, our algorithm along with Louvain behave differently. Meaning that as the density increases, we can detect communities in networks with a more ambiguous structure. The effect of density growth on Louvain is quite different as only its curvature decreases, meaning that not only it starts to fall on later stages, but also its quality decreases less in the early stages of its fall. The CNM algorithm behaves regardless of the density of the network, as it seems to only take into account the clarity of the community structure in all three networks.

Now we take a look at each algorithm's performance in the three networks individually and compare it to ours. BPA and LPA generally follow the same path. They usually yield near-perfect results in the beginning and are the first algorithms to fall to zero. Both of these networks yields results with worse quality compared to ours, in all three of the networks. To be more specific, not only they reach zero sooner than us, but also they yields equal or lower NMI compared to us on networks that they show a reasonable performance. 
DPA's behavior seems to be a little more complicated as it shows a smooth fall in the first network ($\bar{d}=20$) and a sudden fall in the last one ($\bar{d}=40$). The smooth fall of DPA enables it to surpass our algorithm on very high $\mu$ after it goes below our algorithm during $\mu=0.55-0.75$. This does not hold for the other two networks, as DPA completely reaches zero before we even start to fall. The case is quite clear for the Infomap algorithm, as on all three networks, our algorithm outperforms it both in point-to-point comparison and in landing point. As we mentioned earlier, the Louvain plot has a smooth fall. Due to this characteristic of Louvain, we start to surpass it on early stages, as we yield near-perfect results and the Louvain suffers an amount of decrease in the quality of its retrieved communities. But Louvain catches us in the end in all three of the networks. We will further talk about the shortcoming of our algorithm on very high $\mu$s on all three of the networks. However, you should keep in mind that even though the Louvain keeps on giving a community strucutre, on average, the resemblance between the retrieved communities and the true ones, decrease very fast to around 0.2 in terms of NMI. This means that, although Louvain does not completely miss the essence of the community structure present in a network, but it also does not detect a big portion of it. The CNM algorithm behaves regardless of the density of networks and achieves lower NMI compared to all others.

Finally, to find out the cause of our algorithm's failure in very high $\mu$s, we calculated the number of dissatisfied (remember from section \rom{2}.A where we discussed about dissatisfaction of nodes in a given clustering) nodes on all three of the networks for each $\mu$. The results are shown in Fig.\ref{fig:nmi:unhappy}. Since there were not any dissatisfied node when $\mu$ was below 0.5, we did not plot the corresponding data. As you can see, when $\mu$ reaches the end of its spectrum, the number of unhappy nodes increases exponentially. As a result, the ground truth clustering cannot be an equilibrium state for LPA or our algorithm. This means that without altering the essence of LPA, it will be extremely difficult to detect the planted partitions in networks in the end of $\mu$ range. Besides, it is extremely rare to see a $\mu$ this high in real-world networks any way.

\section{Further Works \& Conclusion}

We discussed how LPA has considerable potential to be employed in massive social networks. To name a few, its ability to be utilized in distributed environments, its intuitive approach to community detection in addition to great time and memory efficiency. We then outlined its major flaw, known as the flood-fill phenomenon. We mentioned how previous works have chosen two different paths to address this issue:
\vspace{-1.5mm}
\begin{itemize}[noitemsep,topsep=0pt,parsep=0pt,partopsep=0pt]
	\item Changing the objective function of LPA.
	\item Controlling the speed of growth for all communities.
\end{itemize}
We then proposed an algorithm, choosing the second option, to control over-propagation of popular labels through choosing a small capacity and gradually increasing it over time. We then experimented our algorithm on several domains, namely sparse real-life networks, a dense real-life network and standard benchmarks with planted partitions. 
Our results showcased the strength of our algorithm in networks with unclear community structure, compared to existing robust algorithms. Since our method tackles the problem of flood-fills in a new way (as far as we know), it can be considered as a framework for others to propose a more efficient enhancement over our method to prevent flood-fills. Also, further research can be done in the area of finding other ways to prevent flood-fills starting from the very first iteration. Also an interesting area might be analyzing the properties of real-world networks in which LPA completely fails. Moreover, our method of preventing flood-fill can be easily incorporated with the majority of LPA extensions, such as BPA and DPA, since it does not change the essence of LPA by any means.

\section*{Acknowledgment}

The authors would like to thank Dr.{\v{S}}ubelj for kindly providing us with the codes of BPA and DPA algorithms.

\bibliographystyle{IEEEtranN}
\bibliography{./clpa}

\end{document}